\documentstyle[epsfig,amsfonts,amssymb,diagram,preprint,tighten,floats,aps]{revtex}

\newcommand{\R}{{\Bbb R}}
\newcommand{\C}{{\Bbb C}}

\def\S{{\cal S}}

\begin{document}

\title{3D Gravity, Point Particles \\ and  \\ Liouville Theory}

\author {Kirill Krasnov}

\address{Physics Department, University of California,\\
Santa Barbara, CA 93106, USA}

\maketitle

\begin{abstract}

This paper elaborates on the bulk/boundary relation between negative 
cosmological constant 3D gravity and Liouville field theory (LFT). 
We develop an interpretation of LFT non-normalizable 
states in terms of particles moving in the bulk. This interpretation is 
suggested by the fact that ``heavy'' vertex operators of LFT create 
conical singularities and thus should correspond to point particles
moving inside AdS. We confirm this expectation by comparing the
(semi-classical approximation to the) LFT two-point function  
with the (appropriately regularized) gravity action evaluated on 
the corresponding metric.

\end{abstract}

\eject

\section{Introduction}
\label{sec:intr}

The fact that Liouville theory describes asymptotic degrees of freedom of
negative cosmological constant 3D gravity has been recognized some time ago
\cite{H-Liouv}. Since then the appearance of LFT as an effective theory
at asymptotic infinity was re-confirmed by many authors, see 
\cite{Kostas,Bautier} and a more recent reference \cite{Spindel}.
However, all these results demonstrate only a relation between
the classical theories: classical gravity in the bulk and classical 
LFT on the boundary. What concerns us in this paper is the relation
between the corresponding quantum theories. Indeed, there is a growing 
evidence, see \cite{Techner}, that Liouville theory does make sense 
quantum mechanically. It is then interesting to ask: exactly how
much of quantum gravity in the bulk, if any, is described by
quantum LFT on the boundary? This question is of interest for 3D
quantum gravity, for if this theory makes sense by itself, without
embedding into string theory, the quantum LFT is a candidate for
its holographic description. This question is also of interest
in the context of string theory, for LFT is a part of the
sigma model that is believed to be dual to the string theory
on AdS${}_3\times S^3\times T^4$. Thus, by studying quantum LFT one might 
be able to learn something new about string theory on AdS${}_3$,
a subject of recent interest.

In this paper we make a step towards 3D quantum gravity interpretation
of quantum LFT. We consider LFT vertex operators. These are local, or, in 
the terminology of \cite{Seiberg}, microscopical operators, which
create states that are non-normalizable because of the change in the topology
these operators introduce. Depending on its conformal dimension, such an
operator creates either a conical singularity or a puncture at the point
where it acts. In this paper we show that vertex operators on the boundary should be
interpreted as describing point particles in the bulk.\footnote{%
The interpretation of boundary configurations containing conical singularities
in terms of point particles in the bulk is rather natural and was mentioned,
e.g., in \cite{Martinec}. What is new in this paper is an attempt to 
establish this relation also at the quantum level, by comparing the
bulk and boundary partition functions.}
To see this, consider a point particle moving inside Euclidean AdS. 
It creates a line of conical singularities, 
which pierces the boundary at two points, and endows the boundary metric
with angle deficits at those points. This is precisely the same boundary configuration
as the one created by two vertex operators. We are thus led to interpreting
one of these operators as creating a particle, and the other as annihilating it
after its evolution in the bulk. The angle deficit created depends
on the conformal dimension of the operator. As we shall see, the 
CFT relation between mass and conformal dimension translates then into
the usual in gravity relation between mass and the angle deficit. To
give more evidence in support of this interpretation we compare the
bulk and boundary partition functions. More precisely, we consider the semi-classical
approximation, in which the bulk and boundary partition functions are dominated
by the corresponding classical solutions, and show that the (appropriately
regulated) bulk action evaluated on Euclidean AdS with a line
of conical singularities is equal to the Liouville theory action evaluated on 
the Liouville field that describes a metric with two conical singularities.

Thus, vertex operators of LFT on the boundary receive the 3D 
interpretation of operators creating point particles inside AdS. This
interpretation is potentially interesting for 3D quantum gravity, for one
can use the results available from LFT to analyze processes involving
point particles, such as, e.g., the black hole creation process of \cite{Matsch}.
On the other hand, this interpretation gives a new perspective on LFT itself,
for it means that the non-normalizable states of LFT that are created by
vertex operators are physically interesting states of the theory and
should receive more attention.

At this point it seems appropriate to mention that the idea to consider 
quantum LFT as a theory of quantum gravity in 3D was a
subject of criticism, in particular by \cite{Martinec}.
The reason for this criticism is as follows. Although by tuning the coupling constant
the central charge of Liouville theory can be made to be equal 
to Brown-Henneaux central charge \cite{Br-Hen}, it is known that if one
only considers the normalizable states in LFT, because of the
mass gap, the effective central charge is rendered to be one.
Since it is the effective central charge that appears in the
formula for the asymptotic density of states, the number of 
states in the quantum LFT is then not enough to account for
the BTZ black hole entropy, see \cite{Carlip} for a good discussion of 
this issue. Then it seems unavoidable \cite{Martinec} that
one needs the whole of string theory to account for the microscopic
origin of the black hole entropy, and that LFT only describes
macroscopic, thermodynamical degrees of freedom of gravity.
However, in view of the interpretation of the non-normalizable states of LFT as
physical states corresponding to the point particles inside
AdS, one may have to reconsider the issue with the 
density of states. Indeed, if one allows the non-normalizable
states, then there is no mass gap in the spectrum, which
might mean that one has to consider the full central charge in
the formula for the density of states. While this remains to be
shown, it is certainly true that the point particle states are
there to be considered as the states contributing to the black
hole entropy, and this might change the state counting. This might
invalidate the point of \cite{Martinec}. We, however, do not attempt to settle
this issue in the present paper. Let us note that non-normalizable modes 
were considered as relevant to the black hole entropy problem in \cite{Myung}.

Thus, in this paper we take --as a working hypothesis-- the viewpoint 
that quantum LFT provides one with a holographic description of quantum
gravity, or some sector of it. 
Although the possibility of such a relation between 
the two quantum theories has been discussed, a precise identification between 
LFT and gravity, namely a relation between the cosmological constant and the 
coupling constant of Liouville theory, does not seem to have appeared
in the literature. The correspondence we propose is as follows. The
3D Einstein-Hilbert action is:
\begin{equation}
\label{action-gr}
S_{gr} = {1\over 16\pi G} \int_{\cal M} d^3x \sqrt{g} (R-2\Lambda),
\end{equation}
where $G$ is Newton constant, the integral is taken over
the 3D manifold $\cal M$ that we consider gravity on, 
$\Lambda$ is the cosmological constant,
which we assume to be negative, $g$ is the determinant of the
metric, which we consider to be of Euclidean signature, $R$ is the
trace of Ricci. The Liouville theory action is:
\begin{equation}
\label{action-L}
A_L[\phi] = 
\int d^2x \left[ {1\over 4\pi} (\partial_a \phi)^2 + \mu e^{2b\phi} \right].
\end{equation}
Here $\phi$ is the Liouville field, $b$ is the coupling constant of the
theory and $\mu$ is a constant of the dimension of $1/({\rm length})^2$,
which sets a scale for the theory. We leave the region over which the integral 
is taken unspecified for now. All physical quantities in LFT depend not on $b$, 
but on the quantity
\begin{equation}
\label{Q}
Q = b + {1\over b}.
\end{equation}
The identification we propose is:
\begin{equation}
\label{rel}
Q^2 = {l\over 4G\hbar} = {l\over 4 l_p},
\end{equation}
where $l = 1/\sqrt{-\Lambda}$ is the curvature radius of AdS. The regime when
the semi-classical approximation in LFT is valid (small $b$) corresponds,
according to (\ref{rel}) to large curvature radii, and, thus, is also
the regime when the semi-classical approximation in gravity should be valid.
In this regime, the Liouville central charge:
\begin{equation}
\label{c}
c_L = 1 + 6Q^2
\end{equation} 
matches the Brown-Henneaux central charge \cite{Br-Hen} $3l/2G\hbar$.
Interestingly, the so-called strong coupling region in LFT,
$1< c_L < 25$, of which very little is known, corresponds in view of
(\ref{rel}), to the Planckian regime $l < 16 l_p$ on the gravity side.
Another interesting fact is that, unlike in AdS${}_5$/CFT${}_4$ correspondence,
where there is only one limit on the CFT side that corresponds to the
semi-classical limit on the gravity side, there are two different
limits one can reach this regime in our example. Because of the
famous Liouville theory duality $b\to 1/b$, both small and large
coupling regimes in LFT correspond to the semi-classical
regime on the gravity side.

The paper is organized as follows. Section \ref{sec:prel} reviews
some basic facts about quantum LFT. Section \ref{sec:part} gives
the point particle interpretation of LFT vertex operators. We conclude with 
a discussion of the results obtained.

We should emphasize that, although the relation between 3D gravity
and Liouville theory can be discussed for the Lorentzian signature as 
well, this paper treats only the Euclidean case. To interpret the Euclidean 
results we obtain in terms of the real world Lorentzian signature 
gravity one has to adopt a version of the analytic continuation
procedure. Although this question is important, it is not
considered in this paper.

\section{Preliminaries: quantum Liouville theory}
\label{sec:prel}

In this section we review some well-known facts about Liouville theory.
Our main source is \cite{Zamolo}, which the reader is referred to
for more details. Our conventions and notations on LFT are the same as 
in this reference.

Liouville theory is described by the action (\ref{action-L}). Usually
one also adds to the Lagrangian a term proportional to $\phi R$, where 
$R$ is the curvature scalar of a fixed background metric, and 
adjusts the coefficient in front of this term so that the action is
independent of the background. For our purposes, however, it is
more convenient to work in the flat background (this is also the
choice of  \cite{Zamolo}). Then the integral of $\phi R$ translates into 
a bunch of boundary terms, see below. Although by appropriately choosing the
boundary terms one can define LFT on any Riemann surface, see \cite{Takht}, 
for the purposes of this paper it is sufficient to consider the case of a 
sphere. The LFT on a sphere corresponds to field $\phi$ defined on the
whole complex plane with the following asymptotics for $|z|\to\infty$:
\begin{equation}
\label{b-cond}
\phi(z,\bar{z}) = - Q \ln(z\bar{z}) + O(1),
\end{equation}
where $Q$ is given by (\ref{Q}). To make the action (\ref{action-L})
well-defined on such fields, one introduces a large disc $\Omega$ of radius
$R\to\infty$ and adds a boundary term to the action:
\begin{equation}
A_L = 
\int_\Omega d^2x \left[ {1\over 4\pi} (\partial_a \phi)^2 + \mu e^{2b\phi} \right]+
{Q\over \pi R}\int_{\partial\Omega} dl \, \phi + 2Q^2 \ln{R}.
\end{equation}
The last term is needed to make the action finite as $R\to\infty$.

The vertex operators of LFT are:
\begin{equation}
V_\alpha(x) = e^{2\alpha\phi(x)},
\end{equation}
where $x$ is a point on $S^2$. They are primary operators of conformal dimension:
\begin{equation}
\Delta_\alpha = \alpha(Q-\alpha).
\end{equation}

Correlation functions of vertex operators are formally defined
as the following functional integral:
\begin{equation}
\label{corr}
{\cal G}_{\alpha_1,\ldots,\alpha_n}(x_1,\ldots,x_n) =
\int {\cal D}\phi \, V_{\alpha_1}(x_1)\cdots V_{\alpha_n}(x_n)
e^{-A_L[\phi]}.
\end{equation}
The integral must be taken over the fields satisfying the boundary 
condition (\ref{b-cond}).

The scale dependence of correlators is:
\begin{equation}
{\cal G}_{\alpha_1,\ldots,\alpha_n}(x_1,\ldots,x_n) =
(\pi\mu)^{(Q-\sum\alpha_i)/b} F_{\alpha_1,\ldots,\alpha_n}(x_1,\ldots,x_n),
\end{equation}
where $F_{\alpha_1,\ldots,\alpha_n}(x_1,\ldots,x_n)$ is independent
of the scale $\mu$. Instead of correlation functions at
fixed scale $\mu$ we shall often consider the fixed area
correlation functions ${\cal G}^{(A)}_{\alpha_1,\ldots,\alpha_n}(x_1,\ldots,x_n)$, 
given by the functional integral over fields with 
\begin{equation}
A = \int dx^2\, e^{2b\phi}
\end{equation}
fixed. The fixed area correlators are related to (\ref{corr}) by
\begin{equation}
{\cal G}_{\alpha_1,\ldots,\alpha_n}(x_1,\ldots,x_n) =
\int_0^\infty {\cal G}^{(A)}_{\alpha_1,\ldots,\alpha_n}(x_1,\ldots,x_n) 
e^{-\mu A} {dA\over A}\, ,
\end{equation}
so that
\begin{equation}
{\cal G}^{(A)}_{\alpha_1,\ldots,\alpha_n}(x_1,\ldots,x_n) = 
\left( {A\over\pi} \right)^{(\sum\alpha_i - Q)/b} 
{F_{\alpha_1,\ldots,\alpha_n}(x_1,\ldots,x_n)\over
\Gamma((Q-\sum\alpha_i)/b)}\,.
\end{equation}

The spectrum of LFT consists of the states created by $V_\alpha$ with 
\begin{equation}
\label{spec}
\alpha = {Q\over2} + iP.
\end{equation}
These are the normalizable states. One can also consider the ``states''
created by $V_\alpha$ with $0<\alpha<Q/2$. These operators create
conical singularities and thus correspond to non-normalizable states.

Classical LFT, which is that with the interaction term being
$e^\phi$, appears as the semi-classical limit of the theory
described by the action (\ref{action-L}). The semi-classical
limit of LFT corresponds to $b\to 0$. One then introduces a
new field:
\begin{equation}
\varphi = 2b\phi,
\end{equation}
which becomes the classical Liouville field in this limit. 
The ``quantum'' action (\ref{action-L}) is then:
\begin{equation}
A_L[\phi] = {1\over b^2} S_{Liouv}[\varphi],
\end{equation}
where
\begin{equation}
\label{action-class}
S_{Liouv}[\varphi] = {1\over 8\pi} \int d^2x \left[
{1\over2} (\partial_a \varphi)^2 + 8\pi\mu b^2 e^\varphi \right]
\end{equation}
is the classical Liouville action. There are also some boundary terms to 
be added to this action, see below. Varying the classical action with 
respect to $\varphi$ one finds that locally $\varphi$ satisfies the 
classical Liouville equation:
\begin{equation}
\label{Liouv-eq}
\Delta \varphi = 8\pi\mu b^2 e^\varphi.
\end{equation}
Then the metric $ds^2 = e^\varphi |dz|^2$ is a metric of constant
negative curvature $-8\pi\mu b^2$. The importance of the classical
Liouville action lies in the fact that correlation
functions of vertex operators are, in the semi-classical limit,
dominated by $S_{Liouv}$ evaluated on the corresponding solution
to the Liouville equation. Let us consider as an example the case of
``heavy'' vertex operators, which is relevant for this paper. 
Let us take $\alpha_i = \eta_i/b$ with $\eta_i$ of order $O(1)$,
and consider the case $\sum\eta_i < 1$ so that there is no
solution to (\ref{Liouv-eq}) with negative curvature.  The relevant
solution is that with positive curvature. One has
to impose the area constraint
\begin{equation}
A = \int dx^2\, e^\varphi
\end{equation}
and consider the positive curvature Liouville equation
\begin{equation}
\Delta \varphi = - {8\pi\over A} \left( 1 - \sum\eta_i \right) e^\varphi
\end{equation}
with the field $\varphi$ satisfying the following boundary conditions:
\begin{eqnarray}\nonumber
\varphi(z,\bar{z}) = - 2\ln{|z|^2} + O(1) \qquad {\rm at}\quad |z|\to\infty\\
\label{asympt}
\varphi(z,\bar{z}) = - 2\eta_i \ln{|z-x_i|^2} +O(1) \qquad {\rm at}\quad z\to x_i.
\end{eqnarray}
The fixed area correlation functions are then dominated by the fixed area
Liouville action:
\begin{equation}
{\cal G}^{(A)}_{\alpha_1,\ldots,\alpha_n}(x_1,\ldots,x_n) \sim
\exp{\left( -{1\over b^2} S_{Liouv}^{(A)} \right)}.
\end{equation}
Here the fixed area Liouville action $S_{Liouv}^{(A)}$ is the
one without the interaction term, and contains 
boundary terms around singularities at $x_i$:
\begin{equation}
\label{action-A}
S_{Liouv}^{(A)}[\varphi] = {1\over 8\pi} \int_\Omega dx^2 \left[ {1\over 2} 
(\partial_a\varphi)^2 \right] + \phi_\infty + 2\ln{R} - 
\sum_i \left ( \eta_i \varphi_i + 2\eta_i^2 \ln{\epsilon_i} \right).
\end{equation}
Here $\Omega$ is a disc of radius $R$ with small discs of radii
$\epsilon_i$ cut out around each of the singularities at $x_i$, and
\begin{equation}
\varphi_i = {1\over 2\pi\epsilon_i} \int_{\partial\Omega_i} dl \, \varphi.
\end{equation}

As an example, let us give the fixed area action for the case 
of two vertex operators, each of
$\alpha = \eta/b$. For simplicity, we insert the operators at $z=0,\infty$.
The classical Liouville field in this case is:
\begin{equation}
\label{Liouv-field}
e^\varphi = {A\over \pi (1-2\eta)} {a^2 (z\bar{z})^{a-1}\over (1+(z\bar{z})^a)^2} ,
\end{equation}
where $a = 1-2\eta$. In the next section we shall see how this field is
obtained. One can then evaluate the fixed area action (\ref{action-A}) 
on this field. The resulting fixed area action for two vertex operators is:
\begin{equation}
\label{two-point}
S_{Liouv}^{(A)} = (1-2\eta)\left( \ln{A\over\pi} + \ln(1-2\eta) - 1 \right).
\end{equation}
In the next section we compare this action, or, more precisely, a related
quantity, to the free energy of AdS metric with a line of conical
singularity.

\section{Particle interpretation}
\label{sec:part}

We would now like to describe a point particle interpretation of the
LFT vertex operators $V_\alpha$ with $0<\alpha<Q/2$. Let us, as before,
when analyzing the semi-classical approximation of LFT, introduce a
parameter $\eta: \alpha=\eta/b$. This parameter measures the angle deficits
of the metric described by the classical Liouville field. To see this,
we must rewrite the vertex operator $V_\alpha$ as:
\begin{equation}\nonumber
V_\alpha = e^{2b\phi\,\eta/b^2} = e^{\varphi\eta/b^2}.
\end{equation}
When $b\to 0$, the correlation functions are dominated by the
extremum of the classical Liouville action (\ref{action-class})
with sources $\sum_i \eta\varphi(x_i)$. This then introduces
$\delta$-function type singularities on the right hand side
of (\ref{Liouv-eq}):
\begin{equation}
\Delta \varphi = 8\pi\mu b^2 e^\varphi - \sum_i \eta_i \delta_{x_i}.
\end{equation}
Recalling that $\Delta\varphi = - \sqrt{g} R$, where $g$ is the 
determinant of the metric
$ds^2 = e^\varphi |dz|^2$ and $R$ is its curvature, we see that the angle
deficits at $x_i$ are equal to $4\pi\eta_i$. 

Let us recall now that the analysis \cite{Br-Hen} of asymptotic symmetries 
of $\Lambda<0$ 3D gravity gives the following relation between mass and
conformal dimension:
\begin{equation}
\label{mass}
M l = \Delta+\bar{\Delta}.
\end{equation}
It is exactly this relation between $M$ and $\Delta$ (plus an analogous relation
for the angular momentum) that needs to be
inserted into the Cardy formula for the CFT asymptotic density of states
to get the correct Bekenstein-Hawking entropy for BTZ black hole,
see \cite{Strom}. For vertex operators under consideration the 
conformal dimension is real, which means that particles they are
to describe are non-rotating. Let us consider the case of ``not very heavy''
vertex operators: $\eta \ll 1$. Then $\Delta_\alpha\approx Q\alpha\approx \eta/b^2$.
Recalling the relation (\ref{rel}) between $l$ and $Q$, we see that, in this
regime, the mass is given by $M = \eta/2G$. This means that the angle 
deficit is $8\pi GM$, which is correct gravity relation between 
mass and the angle deficit. We thus see that it holds for our vertex
operators, provided the parameter $\eta$ is small. This is the first
check of the particle interpretation of vertex operators.

\begin{figure}
\centerline{\hbox{\epsfig{figure=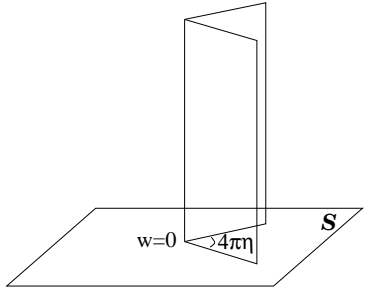,height=1.5in}}}
\bigskip
\caption{The AdS space with a line of conical singularities can be
described as AdS with a wedge removed.}
\label{fig:1}
\end{figure}

To perform a more serious check we need to compare the partition functions
on the bulk and boundary sides. Although the two-point function on the
LFT side is known for any value of the coupling, see \cite{Dorn,Zamolo}, 
there is no quantum gravity calculation of the point particle partition function. 
The best we can do is to compare the partition functions in the regime 
$l\gg l_p$ when one can use the semi-classical approximation in the bulk. 
This is the regime $b\ll 1$ on the LFT side, and, thus, one also uses 
the semi-classical approximation here. The  semi-classical 
approximation to the fixed area two-point function in LFT
is given by $\exp{-S_{Liouv}^{(A)}/b^2}$ with the fixed area action
given above by (\ref{two-point}). We have to compare this
quantity to the $\exp{-S_{gr}}$ evaluated on the AdS metric
with a line of conical singularities. The metric
describing a point particle moving in a constant curvature space
was first obtained in \cite{Deser}. 

Let us describe this geometry in a language that can be easily
generalized to more complicated situations, e.g., with more
than one particle. The metric is obtained as a discrete identification of the
AdS space with respect to a discrete subgroup $\Sigma$ of the
isometry group, which in our case is ${\rm SL}(2,\C)$. To obtain
a single line of conical singularities inside AdS we have to take
$\Sigma$ to be generated by a single elliptic element. Then 
the line of conical singularities is the line of fixed points of
the transformation generated by this element. Action of an 
isometry inside AdS generates a conformal transformation on the
boundary. Any such transformation has two fixed points (possibly
coinciding), and in our case these are the points on the
boundary where it is intersected by the particle worldline.
The quotient of the boundary with respect to $\Sigma$ is,
in general, a Riemann surface, and, in our case, a sphere
with two conical singularities. Let us denote the boundary by
$\S$, and the quotient $\S/\Sigma$ by $X$. Properties of the
projection map $\pi_\Sigma\to X$ are very important. It is the
knowledge of this map that allows one to find the corresponding
Liouville field, which in our example is given by (\ref{Liouv-field}).
Let us now see how this technology works.

It is most convenient for our purposes to use the upper half-space
model of AdS. The hyperbolic metric in this model is:
\begin{equation}
ds^2 = {l^2\over\xi^2} \left( d\xi^2 + |dw|^2 \right),
\end{equation}
where $\xi$ is the coordinate that runs orthogonally to the
boundary $\S$ located at $\xi=0$, and $w$ is a homomorphic coordinate
on $\S$. Note that $\S$ is the whole extended complex plane,
for the point at infinity also belongs to it. In this model
geodesics are half-circles (or straight lines) orthogonal to the 
boundary. By a conformal transformation we can always put
one of the conical singularities (=fixed points) at
$w=0$ and the other at $w=\infty$. Then the particle's worldline
is a straight line orthogonal to the boundary. 
Let $\Sigma$ be generated by a rotation on an angle $2\pi a$ around this 
line. The angle deficit created is then $2\pi (1-a)=4\pi\eta$, see
Fig. \ref{fig:1}. 

\begin{figure}
\centerline{\hbox{\epsfig{figure=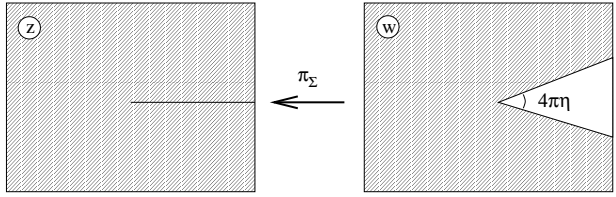,height=1in}}}
\bigskip
\caption{Complex planes $z$ and $w$. The map from
$w$ to $z$ is a covering, which is branched at fixed points
of elliptic generators, in our case at $0,\infty$.}
\label{fig:2}
\end{figure}

The quotient $X$ of $\S$ by $\Sigma$ is again a complex plane,
with two singular points. Let us introduce a homomorphic coordinate $z$
on $X$. The projection map $\pi_\Sigma: \S\to X$ is easy to determine:
\begin{equation}
\label{map}
\pi_\Sigma : w \to z = w^{{1\over a}}.
\end{equation}
Note that it is finitely branched only for integer $a$.
As we have said, the knowledge of this map allows one to 
determine the Liouville field. In our case we would like
to have a Liouville field on the $z$-plane that describes
a constant positive curvature metric of total area $A$.
We get such a field from the sphere metric on the $w$-plane.
This metric is given by:
\begin{equation}
ds^2 = {A\over \pi (1-2\eta) } {|dw|^2 \over (1+ w\bar{w})^2},
\end{equation}
where we have divided $A$ by the factor $(1-2\eta)$ to correct
for the fact that only a portion of the $w$-plane is mapped
to the $z$-plane. Using the inverse of the map (\ref{map}) it
is easy to see that the Liouville field on the $z$-plane is
indeed given by (\ref{Liouv-field}). Comparing the
asymptotics of this field for $z\to 0$ with that given
in (\ref{asympt}) one finds that $(1-a)$ must indeed
be identified with $2\eta$, and, thus, $4\pi\eta$ is the
angle deficit. The simple analysis of this paragraph
can be generalized to much more complex situations. 
It is true in general that the knowledge of the projection
map $\pi_\Sigma$, or rather of its inverse, which is called the
uniformization map, is equivalent to the knowledge of the
corresponding Liouville field. A generalization of the construction 
described dates back to Poincare, who used it as an approach to the problem
of uniformization of Riemann surfaces.

Having described the quotient geometry, let us evaluate the
gravity action on the corresponding metric. In 3D Einstein-Hilbert
action reduces on shell to the volume of the corresponding space:
\begin{equation}
S_{gr} = - {V\over 4\pi G l^2}.
\end{equation}
For spaces that are non-compact, as in our case, this volume diverges, 
and we need to employ some kind of regularization procedure. In asymptotically
AdS spaces, where the volume grows as $A l/2$, where $A$ is the
boundary area, one has a natural regularization by subtracting $A l/2$. 
Thus, the idea is to introduce a family of regularizing surfaces, labelled
by a parameter $\epsilon$, which approach the boundary
as $\epsilon\to 0$. One should then calculate the volume $V_\epsilon$ above a
surface of fixed $\epsilon$, subtract $A_\epsilon l/2$, and take the
limit $\epsilon\to 0$. As we shall see, the area subtraction does not
kill all of the divergence: there is still a logarithmic divergence to
be taken care of. However, as it was discussed, e.g., in \cite{Anomaly},
this divergence does have a CFT interpretation,
namely that of a scale dependence of the partition functions due to
the conformal anomaly.

Let us first see how this procedure works on the simplest case of AdS
space itself. The modification to incorporate the deficit angle is
then straightforward. Let us first do the calculation in the unit ball model,
where we don't have to deal with subtleties coming from the point at infinity.
The AdS metric in the unit ball model, in spherical coordinates is:
\begin{equation}
ds^2 = {4l^2\over (1-r^2)^2} ( dr^2 + r^2\sin^2\theta d\phi^2 + r^2 d\theta^2 ).
\end{equation}
The volume inside a sphere of radius $r=R$ is:
\begin{equation}
\label{b1}
V_R = 4\pi l^3 \left[ {R(1+R^2)\over (1-R^2)^2} - {1\over 2}
\ln{{1+R\over 1-R}} \right].
\end{equation}
The area of this sphere is:
\begin{equation}
\label{b2}
A_R = 4\pi l^2 {4R^2\over (1-R^2)^2}.
\end{equation}
The regularized gravity action is then:
\begin{equation}
\label{b3}
S_{gr} = - {1\over 4\pi G l^2} \left[ V_R - {A_R l\over 2} \right] =
- {l\over 4G} \left[ {4R\over (1+R)^2} - 2\ln{{1+R\over 1-R}} \right].
\end{equation}
Introducing $\epsilon: R=1-\epsilon$, and using the relation
(\ref{rel}) for $b \ll 1$, we get:
\begin{equation}
\label{4}
S_{gr} = {1\over b^2} \left[ -1 + \ln{{A_\epsilon\over \pi l^2}} + O(\epsilon) \right].
\end{equation}
The expression in square brackets, in the limit $\epsilon\to 0$, coincides
with the fixed area Liouville action for the sphere, which can be
obtained from (\ref{two-point}) for $\eta=0$. 

It is now straightforward to modify the result (\ref{4})
to incorporate the angle deficit. First,  when integrating over
$\phi$ to get (\ref{b1}), (\ref{b2}), we have to integrate not
over the whole $2\pi$ but over $2\pi(1-2\eta)$. This modifies
(\ref{b1}), (\ref{b2}) by multiplying them by $(1-2\eta)$.
Second, to get
(\ref{4}) from (\ref{b3}), we have to express $2\ln{(1+R)/(1-R)}$ in
terms of $A_\epsilon$. This introduces another modification.
We now have:
\begin{equation}
2\ln{{1+R\over 1-R}} = \ln{A_\epsilon\over \pi(1-2\eta) l^2} + O(\epsilon).
\end{equation}
Thus, the final result for the regularized gravity action is:
\begin{equation}
\label{final}
S_{gr} = {1\over b^2} (1-2\eta) \left[ -1 + \ln{{A_\epsilon\over \pi l^2}}  
- \ln{(1-2\eta)} + O(\epsilon) \right].
\end{equation}
This is almost the same as (\ref{two-point}), except for the sign in
front of the last term in square brackets. This seeming discrepancy 
comes from the fact that we are comparing the gravity action (\ref{final})
to the wrong Liouville action. Namely, there are two different Liouville actions
to consider: one evaluated on the Liouville field on the $z$-plane,
this action is given by (\ref{two-point}), and another action
calculated for the Liouville field on the $w$-plane. To go between
the two planes one uses a conformal transformation, see Fig. \ref{fig:2}.
Thus, the two actions are not the same, and differ by the 
action evaluated on the Liouville field that describes this 
conformal transformation. It is not hard to show that this
difference between the Liouville actions on the $z$ and $w$-planes
is exactly $(1-2\eta)\ln{(1-2\eta)}$. Thus, the (regularized) gravity action
evaluated on AdS metric with a line of conical singularities {\it does
agree} with the semi-classical approximation to the LFT 
two-point function, when the later is computed as the Liouville 
action on the $w$-plane. 

We would now like to present another way to perform the calculation
that leads to (\ref{final}). Instead of using the unit ball model,
we shall now use the upper half-space picture. Although this other
calculation may seem more awkward, one {\it does have} to use this
type of calculation in more complicated situations, e.g., when the
boundary is a Riemann surface, as was considered in \cite{Riemann},
or when we have more than two punctures. We would like to present this
other way of getting the same result, first, because it is
easily generalizable to other situations, and, second, because
one can clearly see how the Liouville theory (on the $w$-plane)
appears.

\begin{figure}
\centerline{\hbox{\epsfig{figure=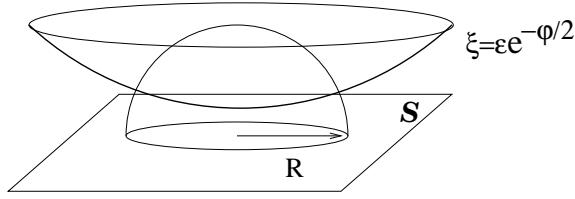,height=1in}}}
\bigskip
\caption{Surfaces of the upper-half space calculation.}
\label{fig:3}
\end{figure}

Thus, let us work in the upper half-space. We need to choose a family of 
regularizing surfaces. A natural choice is:
\begin{equation}
\label{surface}
\xi(w,\bar{w}) = \epsilon e^{-\varphi(w,\bar{w})/2},
\end{equation}
where $\varphi$ is the Liouville field corresponding to the sphere:
\begin{equation}
\label{2}
e^{\varphi(w,\bar{w})} = {A\over \pi (1+w\bar{w})^2}.
\end{equation}
Here 
\begin{equation}
A=\int_{\S} dx^2 e^\varphi.
\end{equation}
For small $\epsilon$, and for sufficiently small $|w|$, surfaces
(\ref{surface}) are just the spheres $r=1-\epsilon\sqrt{4\pi/A}$
of the unit ball model. However, unlike the surfaces $r=const$,
the surfaces (\ref{surface}) touch the boundary $\S$ at $w=\infty$.
Because of this, the volume $V_\epsilon$ above a surface (\ref{surface})
diverges. Let us introduce an additional regularization by a half-sphere
of a radius $R$, see Fig. \ref{fig:3}.

Let us now calculate the volume above a surface (\ref{surface}) and inside
the half-sphere. It is given by:
\begin{equation}
\label{1}
l^3 \int dx^2 \left ( {1\over 2\xi_{min}^2} - {1\over 2\xi_{max}^2} \right).
\end{equation}
Here $\xi_{min}^2(x) = \epsilon^2\exp{-\varphi(x)}$ is the $\xi$
coordinate squared of the $\epsilon$-surface, and 
$\xi_{max}^2(x)=R^2-|x|^2$ is that for the half-sphere. The integral
is taken over the region $x: \xi_{min}(x) \leq \xi_{max}(x)$. The
integral of the second term in (\ref{1}) can be taken. Then (\ref{1})
becomes:
\begin{equation}
l^3 \int dx^2 \, {1\over 2\epsilon^2} e^\varphi - l^3
\left( \pi \ln{{R\over\epsilon}} + {1\over 4R} \int dl\,\varphi \right) + 
O(\epsilon),
\end{equation}
where the last integral is taken along the circle $|x|=R$. We also
need to know the area of the $\epsilon$-plane. It is given by:
\begin{equation}
l^2 \int dx^2 \left( {1\over\epsilon^2} e^\phi + {1\over 8} (\partial_a\varphi)^2
+O(\epsilon) \right),
\end{equation}
with integral taken over the same region as in (\ref{1}). We get the
regularized action by combining the volume and area:
\begin{equation}
\label{3}
S_R = {1\over b^2} \left( {1\over 8\pi} \int dx^2\,(\partial_a\varphi)^2
+ \ln{{R\over\epsilon}} + {1\over 4\pi R} \int dl\,\varphi +O(\epsilon) \right).
\end{equation}
Note that we have not yet used the expression (\ref{2}) for 
$\varphi$. Note also that the expression in brackets is almost
the constant area Liouville action (\ref{action-A}) for the case 
of no vertex operators. The only
difference is that to get (\ref{action-A}) one has to take the
last two terms in (\ref{3}) twice. This happens because our
surfaces (\ref{surface}) behave at infinity differently 
from $r=const$ surfaces of the unit ball model. Our surfaces
are only ``good'' near the origin of the complex plane. To cure
the situation one can either introduce a ``charge'' at infinity,
or, as we shall do, integrate only over a finite region of the
complex plane $|x|\leq R=1$ and then take the result twice. Thus,
we do the calculation by dividing the AdS in two halfs and calculating
the action for one of them. It is not hard to see that this gives the
right result:
\begin{equation}
\label{5}
S_{gr} = 2S_{R=1} = {1\over b^2}
\left( - 1 + \ln{{A\over\pi}} - \ln\epsilon^2 +O(\epsilon) \right).
\end{equation}
Here we have used the expression (\ref{2}) for $\varphi$ and
took the integral of the ``kinetic'' Liouville term. This is
the same result as we obtained earlier in the unit ball model,
one just has to replace $A_\epsilon/l^2$ in (\ref{4}) by
$A/\epsilon^2$ of (\ref{5}). The above calculation makes it clear, 
see (\ref{3}), how the Liouville action appears. Also, as we already
mentioned, it is this calculation that can be generalized to more complicated
situations, in fact, to the case of the boundary being 
any Riemann surface, see \cite{Riemann}. It is true in general
that, regularizing the gravity action with the family of surfaces
(\ref{surface}), where $\varphi$ is the relevant Liouville field,
one obtains the classical fixed area Liouville action as the
result. A similar derivation of the Liouville action was used in
\cite{D1/D5}, although the authors used a different coordinate
system. 

One can now modify the result (\ref{5}) to incorporate
the angle deficit. This is done similarly to what we did to
get from (\ref{4}) to (\ref{final}). Again, one finds that
the regularized gravity action is given by (\ref{final}).
To compare it with (\ref{two-point}) we have to recall that
there is a rescaling (\ref{map}) to go from $w$ to $z$ plane.
Taking into account this rescaling, one finds that the gravity
and LFT actions agree.

\section{Discussion}
\label{sec:disc}

Thus, the particle interpretation of LFT vertex operators passes two
tests: (i) CFT relation (\ref{mass}) between mass and conformal dimension
becomes for these vertex operators the usual gravity relation between
mass and the angle deficit (for small deficit angles); (ii) partition
functions agree, in the semi-classical approximation $b\ll 1, l\gg l_p$.
As we mentioned in the text, this agreement of the partition functions
holds in a much more general situation than that of a single point
particle. There is a computation, along the lines of our upper half-space
computation of the previous section, which works also for compact
Riemann surfaces, see \cite{Riemann}, and can be extended to the most
general case of Riemann surfaces with conical singularities and punctures.
It shows that the regularized gravity action calculated using the
regularizing surfaces (\ref{surface}) is equal to $1/b^2$ times the
fixed area Liouville action, that is the one without the $e^\varphi$
term\footnote{%
The fact that one gets the fixed area classical action, that is, the
one of the {\it non-interacting} Liouville theory, is as one expects.
Indeed, removing the regulator corresponds to going to a fixed 
point of the renormalization group flow, at which all scales must
disappear. Thus, there cannot be an area term in Liouville action,
for such term would set a scale for the theory. We thank S.\ Solodukhin for
pointing out to us the renormalization group interpretation.}, 
evaluated on the relevant Liouville field. 
Thus, at least at the semi-classical level, the partition functions
for gravity and LFT coincide, provided the identification (\ref{rel})
is made.

This is still, however, a classical result, because to arrive to it
we only had to compare the classical actions. It does not by itself imply
that quantum LFT has something to do with quantum gravity in 3D.
To establish a relation between quantum theories we would have to
calculate some quantity in quantum gravity, not in the semi-classical
approximation, and then compare this quantity to the one calculated on the
CFT side. The lack of a quantum theory of negative cosmological constant
3D gravity prevents us from establishing such a relation. 
Note, however, that the situation here is not much worse than that with AdS/CFT dualities,
for which most of the evidence so far comes from the semi-classical 
checks. In our case we cannot do better because we don't have a 
good theory of $\Lambda<0$ 3D quantum gravity, in string theory this happens because it
is hard to do the full fledged string theory in backgrounds containing
AdS.

There is, however, another line of reasoning, based on works on quantum
dilogarithm \cite{K1,K2}, quantization of Teichmuller spaces \cite{K3,K4,K5}
and \cite{Fock1,Fock2}, and quantum Liouville \cite{Techner,Techner-Harm},
which gives strong additional support to the identification of quantum LFT 
with quantum theory of gravity in 3D. In three dimensions one
can construct a discrete, lattice-type model of quantum gravity. 
Such a model was so far formulated for $\Lambda=0,
\Lambda>0$ Euclidean 3D gravity. In the case $\Lambda=0$ it is
known as Ponzano-Regge model \cite{PR}, the $\Lambda>0$ case
is usually referred to as Turaev-Viro model \cite{TV}. These
are state sum models, for one first chops the space into, e.g., 
tetrahedra, labels edges of these tetrahedra in a certain way, 
and then sums over labellings. As one can show, the state sum is
triangulation independent, that is, it is invariant under changes
of triangulation. It is also known to reproduce the correct
partition functions for a variety of manifolds, where the correct
means that the same partition function can be obtained by more
conventional methods, e.g., using the Chern-Simons path integral.
The key point for us is that the triangulation independence of the
states sum, which is the feature that makes these models well-defined,
can be traced back to the fact that each of these models is based on 
a certain category of group representations. This is the usual ${\rm SO}(3)$
for Ponzano-Regge model and the quantum group ${\rm SU}_q(2)$
for Turaev-Viro model, with $q$ being a root of unity. Thus, the
fact that these models make sense can be traced back to the fact
that the set of irreducible representations of the corresponding
groups closes under the operation of taking the tensor product.
Let us now return to Liouville theory. As the recent work \cite{Techner}
shows, LFT seems to be consistent as a CFT, that is, satisfies the
properties of the conformal bootstrap, because the algebra of
conformal blocks of LFT is based on the algebra of irreducible
representations of a certain quantum group. As the results of
\cite{Techner} indicate, there is a certain series of continuous
irreducible representations of ${\rm SL}_q(2,\R)$, with deformation
parameter $q=\exp{\pi i b^2}$, which is closed under taking
tensor products \cite{Techner-Harm}. This property then guarantees the consistency
of the quantum LFT. What is important for our discussion is that
one could built a discrete state sum model of 3D quantum gravity
out of the same representations that are of key importance for
LFT. This model is guaranteed to be consistent, in the sense that
it is triangulation independent, because of the category properties
satisfied by the set of representations in question. It does not,
however, guaranteed to be finite. Indeed, unlike the sum over a discrete set 
of representations in, e.g., Turaev-Viro model, for this model one would 
have to take an integral over a continuous family of representations. 
Assuming that such integrals can be given sense,
this model would be a good candidate for a quantum theory of 3D
negative cosmological constant gravity. It would be very 
interesting to construct such a bulk model; partial results in this
direction are contained in the works of Kashaev \cite{K1,K2} on knot invariants 
from the quantum dilogarithm. Because the consistency of
both this model and LFT would be based on using the same series of representations,
it is very likely that this model will be related to quantum LFT, the later 
giving a holographic description of the former. A similar discussion of
the relation between state sum models and boundary Liouville theory can be
found in \cite{Martin}.

Accepting a possibility of bulk/boundary relation between $\Lambda<0$ gravity
and Liouville theory, there are many interesting gravity calculations that can be 
done using LFT on the boundary. For example, it would be interesting to analyze
the black hole creation process of \cite{Matsch} in the
boundary theory. Another thing that has to be done is 
to settle the issue with the density of states in the theory.
This could possibly be done by a method analogous to the one
used in \cite{OM2} to check the prediction \cite{OM} for the
spectrum of string theory in AdS${}_3$. We hope to return to
these questions in future publications.

Our last comment is about relation to string theory in AdS${}_3$.
Recall \cite{OM} that the string spectrum in this case consists
of a continuous part, which describes long strings moving off
to infinity, and a discrete part, which is the spectrum of 
fundamental strings inside AdS. It is interesting that the
gravity/LFT relation under discussion seems to give a very
similar picture. The spectrum of LFT consists of a continuous
part (\ref{spec}), which describes the zero mode of the field moving with
a constant momentum to infinity, and another part, corresponding
to vertex operators with $0<\alpha\leq Q/2$. This other part of
the spectrum is interpreted in the present paper as corresponding
to point particles. Note that it becomes discrete if one allows
only the rational angle deficits, a natural choice from the
point of view of functions on the boundary, which then become
finitely branched. This would then be very similar to what one
finds for strings. The continuous part of the spectrum has
the same interpretation in terms of two-dimensional surface 
--worldsheet in string theory and ``boundary'' in our case--
moving off to infinity. However, the discrete part of the 
spectrum corresponds to fundamental strings inside AdS
on the string side, while in our case it would correspond to point
particles moving inside the space. Provided this similarity is not 
superficial, it might lead to a new and interesting relation
between the two so different objects. 

\bigskip

\noindent{\bf Acknowledgements.} I am grateful to L. Freildel for
an important discussion and to G. Horowitz for suggestions on the
first version of the manuscript. This work was supported in part by 
NSF grant PHY95-07065.

\end{document}